\documentclass[aps,showpacs,prep,graphics,twocolumn]{revtex4}
\usepackage{amssymb}
\usepackage[dvips]{graphicx}
\usepackage{amsmath}
\usepackage{bm}
\usepackage{epsfig}

\begin{document}

\title{On gravitational waves generated during inflation under the influence of a dynamical cosmological ``constant".}
\author{$^{1}$L. M. Reyes, \thanks{%
E-mail address: luzreyes@fisica.ufpb.br} $^{2}$Claudia Moreno, \thanks{%
E-mail address: claudia.moreno@cucei.udg.mx } and  $^{2}$Jos\'e Edgar Madriz Aguilar \thanks{%
E-mail address: madriz@mdp.edu.ar} }
\affiliation{$^{1}$ Departamento de F\'isica, DCI, Campus Le\'on, Universidad de Guanajuato, C.P. 37150, Le\'on Guanajuato, M\'exico, and \\
$^2$ Departamento de Matem\'aticas, Centro Universitario de Ciencias Exactas e Ingenier\'{i}as (CUCEI), Universidad de Guadalajara, Av. Revoluci\'on 1500, S.R. 44430, Guadalajara, Jalisco, M\'exico. \\
E-mail: luzreyes@fisica.ugto.mx, claudia.moreno@cucei.udg.mx, madriz@mdp.edu.ar}

\begin{abstract}
We study cosmological gravitational waves generated during inflation under the influence of a decaying cosmological ``constant", in the Transverse-Traceless (TT) gauge. In our approach we consider a non-perturbative contribution of the dynamical cosmological ``constant" to the tensor modes. As an application of the model we study the well-known cases $\Lambda(t)=\sigma H^2$ and $\Lambda(t) =\vartheta H$. The spectrum of gravitational waves for the first case results scale invariant at the end of inflation, whereas for the second case scale invariance is not achieved, leaving this to new proposals of the form: $\Lambda(t)=f(H,H^2)$, in order to include inflation in some $\Lambda(t)$CDM models. We also found that the non-perturbative contributions of $\Lambda(t)$, accelerate the decreasing of the amplitude of gravitational waves during a power-law inflationary stage, by an exponential factor.
\end{abstract}

\maketitle

\vskip .5cm


\section{Introduction}

The study of metric fluctuations in the early universe has been an important topic who has brought a deep insight about the mechanism of structure formation in the universe. We have basically two length scales on which the structures are present:  cosmological and astrophysical. When cosmological length scales are involved, is on inflationary settings where the existence of a background of gravitational waves in the universe is naturally predicted \cite{Staro,Lyth}.\\
The primordial scalar perturbations provided the seeds of large scale structure, which since then had gradually formed the galaxies we observed today, and which have left an imprint in the background of cosmic microwave radiation (CMBR) currently observed. The tensor perturbations have escaped out of the horizon during inflation and thereby they could be completely conserved to form the relic of background gravitational waves studied now through the CMBR. Their amplitude is related with the energy scale of inflation and they are potentially detectable via observations of B-mode polarization in the CMBR, if the energy of inflation is larger than $\sim 3\cdot 10^{15} GeV$ \cite{Kami1,Kami2,Knox}. The basal mechanism for generation of primordial gravitational waves in cosmology has been discussed in \cite{Allen}. In the literature we can find two main formalisms: the coordinate based approach of Bardeen \cite{Bardeen} and the covariant formalism \cite{Ellis}.\\

On the other hand, current observational data based on CMB, Bao and Super-Novaes type Ia, seem to indicate the the universe is now passing for a period of accelerating expansion \cite{Nova1,Nova2}. Several proposals have appeared to explain this late time acceleration in the expansion. The first and most popular is the well known $\Lambda$-CDM model. In this proposal the acceleration is due to the presence of a cosmological constant. However, this model suffers from the called cosmological constant problem \cite{Linder}. The $\Lambda$-CDM models fails in explain the huge difference between the cosmological constant inferred from observational data ($\Lambda\simeq 10^{-52}\, m^{-2}$) and the vacuum energy density resulting from quantum field theory ($\Lambda\simeq 10^{122}\,m^{-2}$), when we associate the cosmological constant with the energy density of va\-cuum. The $\Lambda(t)$CDM class of models are one of the proposals to solve the cosmological constant problem. The main idea in this models is to consider a decaying vacuum energy modeled by a dynamical cosmological ``constant". When $\Lambda(t)$ is introduced, careful must be taken on $\Lambda(t)$ in order to not generate contradictions with nucleosynthesis and structure formation predictions. A number of proposals for a time dependence of $\Lambda(t)$ have been made, however two of the most popular models on this line of reasoning are $\Lambda(t)=\sigma H^{2}$ and $\Lambda(t)\approx m^{3}H$ (\cite{J1} and references there in), where $H$ is the Hubble parameter. When the first case $\Lambda(t)=\sigma H^2$ is considered, the value of $\Lambda$ inferred for the present universe coincides with its observational current value. A vacuum energy decay as $\Lambda(t)\approx m^3 H$, with $m\approx 150$ MeV being the energy scale of the chiral phase transition of QCD, was inspired on Sch$\ddot{u}$tzhold work \cite{Schu1,Schu2}. A cosmological scenario filled with matter and a $\Lambda(t)\approx m^{3}H$, has been recently studied by S. Carneiro, J. S. Alcaniz and collaborators in \cite{J1,C1}. In \cite{C1} it has been shown that the model exhibits the well established features of the standard cosmology. In a more specialized analysis done in \cite{J1}, it has been explored its observational viability by using  three of the most recent SNe Ia compilation sets (Union 2, SDSS and Constitution) along with the current measurements of distance to BAO peaks at $z=0.2$ and $z=0.35$ and the position of the first acoustic peak of the CMB power spectrum. As a result it is reported a very good concordance of $\Lambda(t)$CDM models using $\Lambda(t)\approx m^{3}H$, with the constraints from SDSS and Constitution MLCS2k2 data. In the model is present an initial radiation dominated phase, followed by a matter dominated epoch long enough to permit structure formation, and a final era dominated by the cosmological constant term. However, an early inflationary phase with a dominant component $\Lambda(t)\approx m^{3}H$ is not considered. In this case, we must take into account that in the absence of any kind of matter, during inflation, Einstein equation's point out more to a $\Lambda(t)\approx H^2$ than a $\Lambda(t)\approx H$.\\

Aim of this letter is to study the generation of gravitational waves during inflation, under the presence of a dynamical cosmo\-lo\-gi\-cal constant $\Lambda(t)$. Unlike other approaches for gravitational waves in which the influence of a cosmological constant $\Lambda=\Lambda_{0}$ on the tensor modes is taken just at first order in $\Lambda$, in our approach we consider a non-perturbative dependence of $\Lambda(t)$ in the tensor modes. We work in the Transverse-Traceless (TT) gauge. Thus, we have organized the letter as follows. Section I is devoted to an introduction. Section II is dedicated to the derivation of the dynamical equation for the tensor modes and the mean squared tensor metric fluctuations. In Section III we study some par\-ti\-cu\-lar cases for a decaying $\Lambda(t)$ with relevance in modern cosmology. In section IV we give some final comments.

\section{The tensor modes dynamics}

In order to study the impact of a dynamical cos\-mo\-lo\-gi\-cal ``constant" on the background of gravitational waves generated during inflation, let us start considering the action for the gravitational background plus the free tensor metric fluctuations ${\cal Q}^{i}\,_{j}$ contribution
\begin{equation}\label{a1}
S=\int d^{4}x\sqrt{-g_{(b)}}\left[\frac{R}{16\pi G}+\frac{M_{p}^2}{2}g_{(b)}^{\mu\nu}{\cal Q}^{ij}\,_{;\mu}{\cal Q}_{ij;\nu}\right],
\end{equation}
where $R$ is the Ricci scalar calculated with the background metric $g^{(b)}_{\alpha\beta}$, the semicolon is denoting covariant derivative and $g_{(b)}$ is the determinant of the background metric. The tensor modes ${\cal Q}_{ij}$, besides of its typical dependence on the spacetime coordinates, it is considered an explicit dependence on the dynamical cosmological ``constant" $\Lambda(t)$,  i.e. ${\cal Q}_{ij}={\cal Q}_{ij}(t,\vec{r},\Lambda(t))$.\\

Assuming analyticity of ${\cal Q}_{ij}$ and given their dependence on $\Lambda(t)$ the next taylor expansion around $\Lambda=0$ holds true
\begin{equation}\label{a2}
{\cal Q}_{ij}^{(n)}=\sum_{l=0}^{\infty}\frac{1}{l!}\left[\frac{\partial^{l}}{\partial \Lambda ^l}{\cal Q}_{ij}^{(n)}\right]_{\Lambda = 0}\Lambda^{l}.
\end{equation} 
Inspired on this expansion, we propose the following  ansatz for the tensor metric fluctuations:
\begin{equation}\label{a3}
{\cal Q}_{ij}(x,\Lambda)=h_{ij}(t,\vec{r})e^{\alpha_{0}\Lambda(t)},
\end{equation}
being $h_{ij}(x)$ the usual tensor metric fluctuations in the absence of $\Lambda$ and $\alpha_{0}$ a constant introduced to have the correct units in the exponential i.e. $\alpha_{0}$ has units of $\Lambda^{-1}$. It can be easily seen from the equation (\ref{a3}) that by taking $\Lambda=0$ the standard case without $\Lambda$ is recovered. \\

In the TT-gauge, the perturbed line element for a spatially flat FRW spacetime can be written in the form
\begin{equation}\label{a4}
ds_{pert}^2=dt^{2}-a(t)^{2}(\delta_{ij}+{\cal Q}_{ij})dx^{i}dx^{j},
\end{equation}
where the tensor modes ${\cal Q}_{ij}(t,\vec{r},\Lambda)$ satisfy: $tr({\cal Q}_{ij})={\cal Q}^{i}\,_{j}=0$ and ${\cal Q}^{i}\,_{j;i}=0$. The 3D spatial components of the metric can be written as $g_{ij}=-a^{2}(t)(\delta_{ij}+{\cal Q}_{ij})$ whereas its 3D con\-tra\-va\-riant components can be linearly approximated by $g^{ij}\simeq -a^{-2}(t)(\delta^{ij}-{\cal Q}^{ij})$.\\

Due to the well-known fact that in the TT-gauge only space-space components contribute for tensor fluctuations, the dynamics of the tensor modes is in this case given by linearized Einstein equations 
\begin{equation}\label{a5}
\delta R_{ij}-\frac{1}{2}\,\delta R\, g_{ij}^{(b)}-\left(\frac{1}{2}R^{(b)}-\Lambda(t)\right){\cal Q}_{ij}=0,
\end{equation}
where the label $(b)$ denotes quantities computed with the background metric derived from (\ref{a4}), $\delta R_{ij}$ is the li\-nea\-ri\-zed Ricci tensor and $\delta R$ is the Ricci scalar up to first order in ${\cal Q}_{ij}$.\\

By using (\ref{a4}), after some algebra the equations (\ref{a5}) can be put in the form
\begin{equation}\label{a6}
\ddot{{\cal Q}}^{i}\,_{j}+3H\dot{{\cal Q}}^{i}\,_{j}-\frac{1}{a^2}\nabla^{2}{\cal Q}^{i}\,_{j}+\Lambda {\cal Q}^{i}\,_{j}=0,
\end{equation}
where the dot is denoting derivative with respect the cosmic time. Now by means of (\ref{a3}), writing the equation (\ref{a6}) in terms of the usual tensor fluctuations $h^{i}\,_{j}$ we arrive to 
\begin{eqnarray}
\ddot{h}^{i}_{j}+(3H&+&2\alpha_{0}\dot{\Lambda})\dot{h}^{i}_{j}-\frac{1}{a^2}\nabla^{2}h^{i}_{j}+\left(\alpha_{0}\ddot{\Lambda}+\right.\nonumber \\
&+& \left.\alpha^{2}_{0}\dot{\Lambda}^{2}+3\alpha_{0} H\dot{\Lambda}+\Lambda\right)h^{i}_{j}=0.
\label{a7}
\end{eqnarray}
Introducing the auxiliary field $\chi^{i}\,_{j}(t,\vec{r})$ defined by the transformation
\begin{equation}\label{a8}
h^{i}\,_{j}(t,\vec{r})=e^{-\frac{1}{2}\int (3H+2\alpha_{0}\dot{\Lambda})dt}\chi^{i}\,_{j}(t,\vec{r}),
\end{equation}
the equation (\ref{a7}) can be written as
\begin{equation}\label{a9}
\ddot{\chi}^{i}\,_{j}-\frac{1}{a^2}\nabla^{2}\chi^{i}\,_{j}-\left(\frac{9}{4}H^{2}+\frac{3}{2}\dot{H}-\Lambda\right)\chi^{i}\,_{j}=0.
\end{equation}\\ 
Following the usual quantization process, we implement the Fourier expansion for $\chi^{i}\,_{j}$
\begin{eqnarray}
\chi^{i}_{j}(t,\vec{r})&=& \frac{1}{(2\pi)^{3/2}}\int d^{3}k\sum_{\alpha=+,\times}\,^{(\alpha)}e^{i}_{j}\left[a_{k}^{(\alpha)}e^{i\vec{k}\cdot\vec{r}}\xi_{k}(t) + \right.\nonumber\\
&& +\left.  a_{k}^{(\alpha)}\,^{\dagger}e^{-i\vec{k}\cdot\vec{r}}\xi^{*}_{k}(t)\right],
\label{a10}
\end{eqnarray}
with the creation and annihilation operators $a_{k}^{(\alpha)}\,^{\dagger}$ and $a_{k}^{(\alpha)}$ respectively, satisfying 
\begin{eqnarray}\label{du1}
&& \left[a_{k}^{(\alpha)},a_{k^{\prime}}^{(\alpha^{\prime})}\,^{\dagger}\right]=g^{\alpha\alpha^{\prime}}\delta^{(3)}(\vec{k}-\vec{k}^{\prime}),\nonumber \\
&&\left[a_{k}^{(\alpha)},a_{k^\prime}^{(\alpha^{\prime})}\right]=\left[a_{k}^{(\alpha)}\,^{\dagger},a_{k^{\prime}}^{(\alpha^{\prime})}\,^{\dagger}\right]=0,
\end{eqnarray}
and where the following properties for the polarization tensor $e_{ij}$ hold 
\begin{eqnarray}\label{du2}
&& ^{(\alpha)}e_{ij}=\,^{(\alpha)}e_{ji},\quad k^{i}\,^{(\alpha)}e_{ij}=0,\nonumber\\
&& ^{(\alpha)}e_{ii}=0,\quad ^{(\alpha)}e_{ij}[-\vec{k}]=\,^{(\alpha)}e_{ij}^{*}[\vec{k}].
\end{eqnarray}
We also require the modes $\xi_{k}(t)$ satisfy the commutation relation
\begin{equation}\label{a11}
\left[\xi_{k}(t,\vec{r}),\dot{\xi}_{k}(t,\vec{r}^{\prime})\right]=i\delta^{(3)}\left(\vec{r}-\vec{r}^{\prime}\right),
\end{equation}
which for (\ref{a10}) reduces to
\begin{equation}\label{a12}
\xi_{k}\dot{\xi}_{k}^{*}-\xi_{k}^{*}\dot{\xi}_{k}=i,
\end{equation}
with the asterisk denoting the complex conjugate ope\-ra\-tion. When this condition holds true the modes are normalizable on the UV-sector. \\

In view of (\ref{a9}) and (\ref{a10}), the modes $\xi_{k}$ obey the dyna\-mi\-cal equation
\begin{equation}\label{a13}
\ddot{\xi}_{k}+\left(\frac{k^2}{a^2}-\frac{9}{4}H^{2}-\frac{3}{2}\dot{H}+\Lambda\right)\xi_{k}=0.
\end{equation}
Thus, in principle, for a given $\Lambda(t)$ we can obtain a normalized solution for the $k$-modes $\xi_{k}$, just by solving (\ref{a13}) and imposing the normalization condition (\ref{a12}) to that solution. \\

Once a normalized solution for the $k$-modes is obtained, we are in position to calculate the spectrum for tensor fluctuations
on Super-Hubble scales i.e. on the IR-sector. 

The amplitude of the tensor fluctuations $<h^2>_{IR} = $ $<0|h^{i}\,_{j}h_{i}\,^{j}|0>$ on the IR sector is given in this case by
\begin{equation}\label{a14}
\left<h^2\right>_{IR}=\frac{1}{2\pi^2}e^{-\int (3H+2\alpha\dot{\Lambda})dt}\int_{0}^{\epsilon k_{H}}\frac{dk}{k}k^{3}\left[\xi_{k}\xi_{k}^{*}\right]_{IR},
\end{equation}
where $\epsilon=k_{max}^{IR}/k_{P}\ll 1$ is a dimensionless parameter, being $k_{max}^{IR}=k_{H}(t_i)$ the wave-number related to the Hubble radius at the time $t_i$, which is the time when the modes re-enter to the horizon at the end of inflation. The $k_{P}$ is the Planckian wave-number. For a typical Hubble parameter at the end of inflation $H=0.5\cdot 10^{-9} M_{P}$, values of $\epsilon$ on the range of $10^{-5}$ to $10^{-8}$ correspond to the number of e-foldings $N_{e}=63$.\\

\section{Applications}

Evidently, the amplitude of the tensor modes (\ref{a14}) is depending of the form of $\Lambda(t)$, so in order to study the impact of a dynamical cosmological ``constant" on the background of gravitational waves generated during inflation, it is illustrative to study applications of the model for different cases of $\Lambda(t)$. \\

As we mentioned before, we can find in the literature a wide variety of proposals for $\Lambda(t)$, however, two of the most used are $\Lambda(t)=\sigma H^2$ and $\Lambda(t)=\vartheta H$, with $\sigma$ and $\vartheta$ constant parameters with the appropriated units. According to the Friedmann equation, during inflation $H^{2}\simeq\Lambda(t)/3$, resulting natural the proposal $\Lambda(t)=\sigma H^{2}$, during this epoch. Considering a power-law expansion $a(t)=a_{0}(t/t_0)^p$, with $p\gg 1$ during inflation, the case $\Lambda(t)=\sigma H^{2}$ gives a decaying cosmological ``cons\-tant" $\Lambda(t)\simeq t^{-2}$. On the other hand, when $\Lambda(t)=\vartheta H$, it is easy to show that the Friedmann equation leaves during inflation to a de-Sitter expansion with $a(t)=a_{0}e^{(\vartheta/3)t}$. This scale factor thus corresponds to a cosmological cons\-tant $\Lambda_{0}=\vartheta^{2}/3$. Hence for our purposes, it is convenient to study the particular cases of a cosmological constant $\Lambda_0$ and a decaying $\Lambda(t)$.

\subsection{The case of a decaying $\Lambda(t)$}

Let us considering a power-law inflationary stage, thereby the scale factor is taken to be $a(t)=a_{0}(t/t_0)^p$, with $a_0$ being the present value of the scale factor, $t_0$ the present time, and $p> 1$ a constant parameter compatible with an accelerated expansion. The equation for the modes (\ref{a13}) reduces in this case to
\begin{equation}\label{a15}
\ddot{\xi}_{k}+\left(\frac{k^2}{a_0^2}\,t^{-2p}-\frac{(9/4)p^{2}-(3/2)p}{t^2}-\Lambda(t)\right)\xi_{k}=0.
\end{equation}
For a decaying $\Lambda(t)$ of the form $\Lambda(t)=\varpi/t^2$, with $\varpi$ being a constant parameter, the expression (\ref{a15}) reads
\begin{equation}\label{a16}
\ddot{\xi}_{k}+\left(\frac{k^2}{a_0^2}\,t^{-2p}-\frac{(9/4)p^{2}-(3/2)p+\varpi}{t^2}\right)\xi_{k}=0.
\end{equation}
The general solution for (\ref{a16}) is given by
\begin{eqnarray}
\xi_{k}(t) &=& C_{1}\Gamma(1+\nu)2^{\nu}t^{\frac{1}{2}-(p-1)\nu}z(t)^{-\nu}{\cal J}_{\nu}[z(t)]+\nonumber \\
&+&  C_{2}\Gamma(1-\nu)2^{-\nu}t^{\frac{1}{2}+(p-1)\nu}z(t)^{\nu}{\cal J}_{-\nu}[z(t)],
\label{a17}
\end{eqnarray}
where ${\cal J}_{\nu}$ is the Bessel Function, $z(t)=\frac{k}{a_{0}(p-1)}(t/t_{0})^{1-p}$, and $\nu=\sqrt{1+4\beta}/(2(p-1))$, with $\beta=(9/4)p^2-(3/2)p+\varpi$. Unfortunately, this solution is not normalizable. A normalizable solution of (\ref{a16}) can be obtained when we write (\ref{a16}) in terms of the conformal time $\tau$.  Thus, for the conformal time during inflation  $\tau=[a_0/(t_{0}^{p}(1-p))]t^{1-p}$ the equation (\ref{a16}) gives 
\begin{equation}\label{a18}
\frac{d^{2}\xi_{x}}{d\tau^2}-\frac{p}{1-p}\left(\frac{1}{\tau}\right)\frac{d\xi_{k}}{d\tau}+\left(\kappa^{2}-\frac{\kappa_{0}^{2}}{\tau^2}\right)\xi_{k}=0,
\end{equation}
where $\kappa=(t_{0}^{2p}/a_{0}^2)k$ and $\kappa_{0}^{2}=(1-p)^{-2}[(9/4)p^2-(3/2)p-\varpi]$. The general solution of this equation is given in terms of the first and second kind Hankel functions ${\cal H}_{\mu}^{(1)}$ and ${\cal H}_{\mu}^{(2)}$ in the form
\begin{equation}\label{a19}
\xi_{k}(\tau)=A_{1}\tau^{-\frac{1}{2(p-1)}}{\cal H}_{\mu}^{(1)}[w(\tau)]+A_{2}\tau^{-\frac{1}{2(p-1)}} {\cal H}_{\mu}^{(2)}[w(\tau)]
\end{equation}
being $w(\tau)=k\tau$ and $\mu=[1/(2(p-1))]\sqrt{1+4\kappa_{0}^{2}(p-1)^2}$. The normalized solution to (\ref{a18}) is then
\begin{equation}\label{a20}
\xi_{k}(\tau)=i\sqrt{\frac{\pi}{4}}\left[\frac{(p-1)t_0}{a_{0}^{1/p}}\right]^{-\frac{p}{2(p-1)}}\tau^{-\frac{1}{2(p-1)}}{\cal H}_{\mu}^{(1)}[w(\tau)].
\end{equation}
Now, getting back to the cosmic time, the normalized solution (\ref{a20}) reads
\begin{equation}\label{na1}
\xi_{k}(t)=i\sqrt{\frac{\pi}{4(p-1)}}\,t^{1/2}{\cal H}_{\mu}^{(1)}[w(t)],
\end{equation}
with $w(t)=[kt_{0}^p/(a_0(p-1))t^{1-p}]$. In view of (\ref{na1}), the amplitude for gravitational waves on Super-Hubble scales ($k\gg k_H$), according to the expression (\ref{a14}) gives 
\begin{equation}\label{a21}
\left<h^2\right>_{IR}=\frac{t_{0}^{(3-2\mu)p}}{8\pi^3}\frac{(2a_{0})^{2\mu}}{(p-1)^{1-2\mu}}\Gamma^{2}(\mu)t^{\gamma}e^{-\frac{2\alpha\varpi}{t^2}}(\epsilon k_{H})^{3-2\mu},
\end{equation}
where $\gamma=1-3p-2\mu(1-p)$. Therefore for the spectrum of gravitational waves in this case we obtain
\begin{equation}\label{a22}
{\cal P}_{g}(k)=\left.\frac{t_{0}^{(3-2\mu)p}}{8\pi^3}\frac{(2a_{0})^{2\mu}}{(p-1)^{1-2\mu}}\Gamma^{2}(\mu)t^{\gamma}e^{-\frac{2\alpha\varpi}{t^2}} k^{3-2\mu}\right|_{k=\epsilon k_H}.
\end{equation}
Clearly for $\mu\approx 3/2$ the spectrum is nearly scale invariant. The scale invariance is achieved when the condition
\begin{equation}\label{a23}
\frac{1+3p(3p-2)-4\varpi}{9(p-1)^{2}}=1,
\end{equation} 
is valid. The solution of (\ref{a23}) for $\varpi$ is then $\varpi=3p-2$. The accelerated expansion during inflation $p>1$ leaves to the restriction: $\varpi >1$. Hence, a nearly scale invariant spectrum for gravitational waves under the presence of a $\Lambda(t)=\varpi/t^2$ is perfectly possible if we restrict ourselves to $\varpi >1$.  \\

In the particular case of $\Lambda(t)=\sigma H^{2}$ we would have $\varpi=\sigma p^2$. Thus the condition (\ref{a23}) reduces to
\begin{equation}\label{a24}
\frac{1+(9-4\sigma)p^2-6p}{9(p-1)^2}=1.
\end{equation}
Then for $\sigma$ we have the restriction $\sigma=(3p-2)/p^2$. Hence, for $p>1$ we obtain $0<\sigma\leq 9/8\simeq 1.125$.

\subsection{The case of a Cosmological Constant}

 In the case of a cosmological constant $\Lambda_0$ with a de-Sitter expansion $a(t)=a_{0}e^{H_{0}t}$, with $H_0$ being the cons\-tant value of the Hubble parameter during inflation, the equation for the modes (\ref{a13}) reads
 \begin{equation}\label{b1}
 \ddot{\xi}_{k}+\left(\frac{k^2}{a_{0}^2}e^{-2H_{0}t}-\frac{9}{4}H_{0}^{2}-\Lambda_{0}\right)\xi_{k}=0.
 \end{equation} 
This equation has for solution
 \begin{equation}\label{b2}
 \xi_{k}(t)=D_{1}{\cal H}_{\lambda}^{(1)}[x(t)]+D_{2}{\cal H}_{\lambda}^{(2)}[x(t)],
 \end{equation}
 being ${\cal H}_{\lambda}^{(1,2)}$ the first and second type Hankel functions, $\lambda =[1/(2H_0)]\sqrt{9H_0^2+4\Lambda_0}$ and $x(t)=[k/(a_0 H_0)]e^{-H_{0}t}$. Invoking (\ref{a12}) and taking a Bunch-Davies vacuum condition, the normalized solution for (\ref{b1}) is given by
 \begin{equation}\label{b3}
 \xi_{k}(t)=\frac{i\sqrt{3\pi}}{2\Lambda_{0}^{1/4}}{\cal H}_{\lambda}^{(1)}[x(t)].
 \end{equation}
By using (\ref{b3}) the amplitude for gravitational waves on cosmological scales (\ref{a14}), acquires the form
\begin{equation}\label{b4}
\left<h^2\right>_{IR}=\frac{3}{\pi^3}\frac{2^{-3+2\lambda}}{\sqrt{\Lambda_0}}\frac{\Gamma^2(\lambda)}{(a_0H_{0})^{-2\lambda}}e^{-(3-2\lambda)H_{0}t}(\epsilon k_{H})^{3-2\lambda},
\end{equation}
and the spectrum for gravitational waves results in this case
\begin{equation}\label{b5}
{\cal P}_{g}(k)=\left.\frac{3}{\pi^3}\frac{2^{-3+2\lambda}}{\sqrt{\Lambda_0}}\frac{\Gamma^{2}(\lambda)}{(a_0H_{0}^2)^{-2\lambda}}e^{-(3-2\lambda)H_{0}t}k^{3-2\lambda}\right|_{k=\epsilon k_{H}}.
\end{equation}
It can be easily check from (\ref{b5}) that the spectral index for gravitational waves under the presence of a cosmological constant is $n_{gw}=4-2\lambda$. Thus, in principle, for values of $\Lambda_{0}$ obeying $4\Lambda_0/(9H_{0}^2)\ll 1$, we may have a  nearly scale invariant spectrum $n_{gw}\simeq 1$.\\

A special case that deserves our attention is when $\Lambda_{0}=\vartheta H_{0}$. As we mentioned before, in this case $H_{0}=\vartheta/3$ and $\Lambda_{0}=\vartheta^{2}/3$, and therefore $\lambda =\sqrt{21}/2\simeq 2.2913$. Unfortunately, for this value of $\lambda$ the spectral index becomes $n_{gw}\simeq -0.5826$, which is far away from the scale invariance, thus entering in contradiction with observations, which from WMAP7 $+$ BAO + $H_0$-Mean analysis give  $n_{gw}\simeq 0.963\pm 0.012$ \cite{Smith}. \\

We may interpret this result for $\Lambda(t)=\vartheta H(t)$, as during inflation, at least with respect to the spectrum of gravitational waves, it seems more convenient to use  $\Lambda(t)=\sigma H^{2}(t)$ than $\Lambda(t)=\vartheta H(t)$. However, the fact that the former time dependence $\Lambda(t)=\vartheta H(t)$ works very well for the radiation-dominated, matter-dominated and the present epochs \cite{J1,C1}, suggests that to include inflation we might consider $\Lambda(t)=f(H,H^2)$, in such a way that during inflation the dominant term is $H^2$ and after inflation it becomes sub-dominant in front of $H$. One example of this kind of time dependence for $\Lambda(t)$ would be $\Lambda(t)=(1-\eta_{sr})H^2 +\eta_{sr} H$, where $\eta_{sr}$ is the slow-roll parameter during inflation defined by $\eta_{sr}=-\dot{H}/H^2$. As it is well-known during inflation $\eta_{sr}\ll 1$ while when inflation ends $\eta_{sr}=1$. Thus, we obtain the de\-si\-red do\-mi\-nan\-ce of $H^2$ during inflation and of $H$ for the resting epochs.

\section{final comments}

In this letter we have investigated the impact of a dynamical cosmological constant over the background of gravitational waves generated during inflation. The le\-tter has some differences with respect to the ones we can find in the literature, for example, we incorporate a nonperturbative contribution of the dynamical cosmological constant $\Lambda(t)$ to the tensor modes (see eq.(\ref{a3})). The fact that the contribution is non-perturbative in $\Lambda(t)$ allow us mainly, to calculate new contributions of $\Lambda(t)$ on the spectrum and the mean square amplitude $<h^2>$ on the IR-sector (cosmological scales), at all orders in $\Lambda(t)$. \\

In the literature we can find several proposals for the time dependence of $\Lambda(t)$ (see for example \cite{J1,C1} and references there in). For simplicity we study only two of them, that are peculiar  relations between $\Lambda(t)$ and the Hubble parameter $H(t)$. In the case of $\Lambda(t)=\sigma H^2$, for a power law expansion, we obtain a decaying $\Lambda(t)$ whose contribution during inflation leaves to an scale invariant gravitational spectrum at the end of inflation for $0<\sigma<1.125$. In contrast, for the case of $\Lambda(t)=\vartheta H$ the gravitational spectrum we obtain has a spectral index $n_{gw}\simeq -0.5826$, entering in contradiction with observations. These two facts and the fact that $\Lambda=\vartheta H$ seems to work very well after inflation (see \cite{J1}), may suggest that in order to include inflation, $\Lambda$ may have a dependence in both $H$ and $H^2$, such that during inflation the dominant term would be $\Lambda\simeq H^2$ and after inflation would be $\Lambda\simeq H$. One proposal we can make is  $\Lambda(t)=(1-\eta_{sr})H^2 +\eta_{sr} H$, where $\eta_{sr}$ is the slow-roll parameter during inflation defined by $\eta_{sr}=-\dot{H}/H^2$. Thus during inflation $\eta_{sr}\ll 1$ while when inflation ends $\eta_{sr}=1$, reproducing in this manner  the de\-si\-red do\-mi\-nan\-ce of $H^2$ during inflation and of $H$ for the resting epochs. \\

Another important result derived from our analysis, is that a decaying $\Lambda(t)$ contributes to the mean amplitude of the gravitational waves in a different way than a cosmological constant $\Lambda_0$ does (see eq. (\ref{a14})). In par\-ti\-cu\-lar the presence of $\Lambda(t)$ speeds up the decreasing of the amplitude of the gravitational waves generated during a power-law inflationary stage, by an exponential factor, as it is indicated by (\ref{a22}).

\section{Acknowledgements}

\noindent This work was partly supported by CONACYT Mexico under grant FOMIX-CONACYT 149481  and by Department of Mathematics, CUCEI, Guadalajara University.

\end{document}